\documentclass[twocolumn,showpacs,superscriptaddress,longbibliography,nofootinbib,prl]{revtex4-2}
\usepackage[utf8]{inputenc}
\usepackage{calrsfs}
\usepackage{graphicx}
\usepackage{textcomp}
\usepackage{soul}
\usepackage{amsmath,amssymb}
\usepackage{color}
\usepackage{braket}
\usepackage{xcolor}
\usepackage{float}
\usepackage{physics}
\usepackage{soul}
\usepackage{graphicx}
\usepackage{caption}
\usepackage{amsmath}
\usepackage{amssymb}
\usepackage{bbold}
\usepackage{makecell}
\usepackage{tikz,pgfplots}
\usepackage{pgfplots}
\pgfplotsset{compat=1.18}
\usepgfplotslibrary{fillbetween}
\usetikzlibrary{decorations.pathmorphing, arrows.meta, positioning, calc, shadings, shapes.geometric, shadows.blur, fadings, fit, shapes, backgrounds}
\definecolor{gainRed}{RGB}{215, 48, 39}
\definecolor{lossBlue}{RGB}{69, 117, 180}
\definecolor{deepSpace}{RGB}{40, 40, 50}
\definecolor{gridLine}{RGB}{200, 200, 220}
\usepackage{mathtools}
\usepackage{multirow}
\usepackage{subcaption}
\usepackage{url}
\usepackage[colorlinks=true, urlcolor=blue, citecolor=blue, linkcolor=blue]{hyperref}
\usepackage{etoolbox} 

\newtoggle{issupplement}
\togglefalse{issupplement}

\begin{document}
\title{Black-Hole Microstate Hair Requires a Horizon Measurement}
\author{Sudhanva Joshi }
\email[]{sudhanvajoshi.rs.phy24@itbhu.ac.in}
\affiliation{Department of Physics, Indian Institute of Technology (Banaras Hindu University), Varanasi - 221005, India}

\author{Sunil Kumar Mishra}
\email[]{sunilkm.app@iitbhu.ac.in}
\affiliation{Department of Physics, Indian Institute of Technology (Banaras Hindu University), Varanasi - 221005, India}

\begin{abstract}

Can a smooth horizon hide its microstates? At fixed classical and superselected data, the exterior distinguishability $D_{\max}$ of same-sector infalling states and the departure $\varepsilon$ from isometric infall obey $D_{\max}^{2}+(1-2\varepsilon)^{2}\leq 1$, and every point is realized. Below maximal disruption, the boundary is reached only by which-state measurements: the distinguished pair crosses undisturbed while the exterior records which one fell in. The least a horizon can do to have microstate hair is measure what falls in. Classically hairy black holes are labeled backgrounds, not microstate hair.

\end{abstract}

\maketitle
\textit{Introduction.---}For electrovacuum general relativity, the uniqueness theorems guarantee that a stationary black hole is characterized completely by its mass, charge, and angular momentum \cite{israel1967event,carter1971axisymmetric,robinson1975uniqueness,chrusciel2012stationary}. Beyond electrovacuum, this fails classically: colored black holes in Einstein--Yang--Mills theory \cite{bizon1990colored,volkov1999gravitating}, scalar-haired solutions \cite{herdeiro2015asymptotically}, and electroweak coronas around magnetic black holes \cite{maldacena2021comments} possess regular horizons together with exterior structure beyond $(M, Q, J)$. Discrete gauge charges furnish hair with no classical field at all, detectable only by Aharonov--Bohm interference \cite{krauss1989discrete,coleman1992quantum}. Classical and topological hair are real. The sharp question is therefore not whether black holes can have hair, but whether a \emph{smooth} horizon can hide quantum \emph{microstates}: can two infalling states carrying identical classical and superselected data be distinguished from outside, and at what cost?

Black-hole complementarity answers the first half qualitatively \cite{susskind1993stretched}: no observer sees the infalling state both inside and out, so a horizon that preserves what falls through it should not broadcast what fell. The second half has been missing. How much smoothness must a horizon give up to exhibit microstate hair, is that price attainable, and what does a horizon that pays exactly that price physically do?

We answer all three. Writing $\varepsilon$ for the departure of infall from a perfect isometry, and $\varepsilon=0$ being exact smoothness. $D_{\max}$ is the greatest exterior distinguishability of same-sector infalling states. We then show that the pairs realizable by horizon-crossing channels are exactly those with
\begin{equation}\label{eq:disk}
D_{\max}^{2}+(1-2\varepsilon)^{2}\;\leq\;1\,,\qquad \varepsilon\leq\tfrac{1}{2}\,,
\end{equation}
that the boundary of this region is achieved rather than merely approached, and that, below maximal disruption, the horizons achieving it are precisely which-state measurements: they transmit the states they distinguish undisturbed while depositing outside a record of which one fell in. The cheapest horizon that exhibits microstate hair is therefore one that measures the infalling state, and perfect discrimination of some same-sector pair, $D_{\max}=1$, costs $\varepsilon\geq 1/2$, which is the operating point of a complete measurement on that pair. Complementarity is thereby promoted from a qualitative principle to an exact classification of which horizon channels are possible, together with a physical characterization of the extremal ones.

The result rests on a trade-off theorem of independent interest in quantum information theory, which we prove in the paper: for any channel whose worst-case pure-input fidelity to an isometry is $1-\varepsilon$, the environment can distinguish two inputs at most at level $2\sqrt{\varepsilon(1-\varepsilon)}$, and this is exactly tight, with the equality class identified. Inverting it gives the smoothness cost of any given hair, and the equality analysis supplies the physical characterization above. The consequences are immediate for several programs. It prices the exterior structure of fuzzballs \cite{mathur2005fuzzball}, converts the soft-hair proposal \cite{hawking2016soft} into a quantitative dichotomy, is parametrically compatible, at nonperturbative $\varepsilon$, with the microstate visibility expected from the eigenstate-thermalization ansatz for simple observables, and separates cleanly from computational censorship \cite{hayden2007black,harlow2013quantum}, since it constrains what information exists outside a smooth horizon irrespective of the difficulty of extracting it.

\textit{Setup.---}Consider a black hole in the semiclassical regime, $S_{\rm BH}=A/(4G_N)\gg 1$, and a quantum system $F$ that falls through its horizon. In any theory with long-range fields or topological gauge structure, the infalling Hilbert space decomposes into sectors,
\begin{equation}\label{eq:sectors}
\mathcal{H}_F\;=\;\bigoplus_{c}\,\mathcal{H}_F^{(c)}\,,
\end{equation}
labeled by the exterior-accessible data $c$ carried by $F$: its conserved charges, and any superselected quantum numbers such as discrete gauge charge \cite{krauss1989discrete,coleman1992quantum}. For continuous labels, the direct sum is understood as a direct integral, or as a decomposition into finite-resolution sectors. The black-hole background $\ket{\Phi_0}$ is fixed throughout, including whatever classical or topological hair it carries \cite{bizon1990colored,herdeiro2015asymptotically}; distinct hairy solutions are distinct backgrounds, defining distinct horizon channels, and are not compared by $D_{\max}$ at all. Cross-sector distinguishability is exterior label information that lies outside the definition of $D_{\max}$; smoothness is assessed separately within each sector. The labels are read by exterior measurement in the classical case and by topological interference in the discrete one. Everything that follows refers to a single fixed nontrivial sector $\mathcal{H}_F^{(c)}$, of dimension $d\geq 2$, and this restriction is dictated by the structure of the problem, rather than adopted for convenience: the ideal infall map below is determined by the classical infall data and is therefore sector-indexed by construction, so the smoothness hypothesis exists only sector-wise. Where cross-sector superpositions are physical, as for mass and angular momentum, a formal blockwise extension cannot remain in the smooth-horizon regime, since the bound forces $\varepsilon\geq 1/2$ on a perfectly distinguishable cross-sector pair; and where a superselection rule forbids cross-sector superpositions, as for electric or discrete gauge charge, the hypothesis cannot be formulated across sectors at all (Supplemental Material \cite{SM}, Sec.~VIII). Quantum microstate hair is exterior distinguishability \emph{within} a sector.

After $F$ crosses the horizon, we use the effective bipartition $\mathcal{H}\simeq\mathcal{H}_I\otimes\mathcal{H}_E$ into interior and exterior. This holds, in the sense made precise in the Supplemental Material, for a horizon thickened at the Planck scale, by the split property of quantum field theory \cite{doplicher1984standard,fewster2015split,fewster2016split}, with gauge invariance implemented by edge modes on the entangling surface \cite{donnelly2016geometric}; gravitational dressing is accommodated through the type~II crossed product \cite{witten2022gravity,chandrasekaran2023large}, and the whole construction is given in the Supplemental Material \cite{SM}, Sec.~VII. Horizon crossing is then a quantum channel: with $U$ the unitary implementing infall on $\mathcal{H}_F^{(c)}\otimes\mathcal{H}_{\rm BH}$ and $\ket{\Phi_0}$ the initial black-hole state, the Stinespring isometry $W\ket{\psi}=U(\ket{\psi}\otimes\ket{\Phi_0})$ defines the interior channel $\mathcal{N}_I(\rho)=\mathrm{Tr}_E[W\rho W^{\dagger}]$, its complement $\mathcal{N}_E(\rho)=\mathrm{Tr}_I[W\rho W^{\dagger}]$, and the exterior states $\rho_E(\psi)=\mathcal{N}_E(\ketbra{\psi})$.

The hair of the horizon is the exterior distinguishability of the same-sector infalling states,
\begin{equation}\label{eq:Dmax}
D_{\max}\;=\;\max_{\ket{\psi},\ket{\psi'}\in\mathcal{H}_F^{(c)}}\;
\tfrac{1}{2}\bigl\lVert\rho_E(\psi)-\rho_E(\psi')\bigr\rVert_1\,,
\end{equation}
which by the Holevo--Helstrom theorem \cite{helstrom1969quantum} is, at equal priors, the maximal bias with which any measurement on everything outside the horizon identifies which state fell in. $D_{\max}=0$ is the microstate no-hair condition; $D_{\max}>0$ is microstate hair.

Smoothness is measured against ideal infall. Let $V:\mathcal{H}_F^{(c)}\to \mathcal{H}_I$ be the ideal semiclassical infall map, the state-independent isometry fixed by the classical trajectory together with the associated unitary propagation, and let $F_I(\psi)=\bra{V\psi}\mathcal{N}_I(\ketbra{\psi}) \ket{V\psi}$ be the interior fidelity, the probability that an interior observer testing for the intact infalling state succeeds. Three assumptions define the smooth-horizon idealization.

\noindent\textbf{(A1) Unitarity.} The global evolution $U$ is unitary.

\noindent\textbf{(A2) Horizon causality.} After crossing, no signal propagates from interior to exterior. This does not enter the proof; it is what guarantees that $\rho_E(\psi)$ exhausts the exterior's access to the infall event, so that $D_{\max}$ measures the totality of exterior data on the semiclassical slice under consideration.

\noindent\textbf{(A3) Interior accessibility.} The interior channel is close to the ideal embedding in diamond norm \cite{nielsen2010quantum}: writing $\varepsilon_{\diamond}\equiv\tfrac{1}{2}\lVert\mathcal{N}_I-V(\cdot)V^{\dagger}\rVert_{\diamond}$ for the departure from ideal infall, the smooth-horizon regime being $\varepsilon_{\diamond}\ll 1$. This is the information-preserving component of smooth infall, stated operationally: no single-use discrimination protocol, including one that exploits entanglement with an external reference, distinguishes actual from ideal infall with success probability exceeding $(1+\varepsilon_{\diamond})/2$ under equal priors. Here, smoothness refers specifically to state-preserving infall on the chosen semiclassical code subspace; it does not by itself constrain tidal forces, local stress-energy, or the detailed geometry of the horizon. Assumption (A3) implies $\varepsilon\equiv 1-\min_{\ket{\psi}}F_I(\psi)\leq\varepsilon_{\diamond}$, and the worst-case fidelity deficit $\varepsilon$ is the only consequence our results use and the parameter of the classification below; every lower bound on $\varepsilon$ therefore applies a fortiori to $\varepsilon_{\diamond}$, and the two coincide on the boundary family (Supplemental Material \cite{SM}, Sec.~I).

Physically, $\varepsilon=0$ is exact smoothness; $\varepsilon$ of order $e^{-\alpha S_{\rm BH}}$ is the benchmark size expected of nonperturbative quantum-gravity corrections; and $\varepsilon=1/2$, as shown below, already permits maximal hair. Figure~\ref{fig:setup} fixes the geometry and the channels.

\begin{figure}[t]
\centering
\begin{tikzpicture}[scale=1.1, every node/.style={font=\small}]

  \draw[thick] (2,-2) -- (4,0) node[midway, below right] {$\mathcal{I}^-$};
  \draw[thick] (4,0) -- (2,2) node[midway, above right] {$\mathcal{I}^+$};
  \draw[thick] (0,0) -- (2,-2) node[midway, below left] {$\mathcal{H}^-$};

  \draw[thick, dashed] (0,0) -- (2,2);
  \node[rotate=45] at (0.8, 1.1) {horizon ($\mathcal{H}^+$)};

  \draw[thick] (0,0) -- (-2,2);

  \draw[thick, decorate, decoration={zigzag, segment length=3mm, amplitude=1mm}] (-2,2) -- (2,2);
  \node[above, yshift=2pt] at (0,2) {singularity};

  \node[right, xshift=2pt] at (4,0) {$i^0$};
  \node[below, yshift=-2pt] at (2,-2) {$i^-$};
  \node[above, yshift=2pt] at (2,2) {$i^+$};

  \node at (2.8, 0) {\Large \textbf{E}};
  \node at (-0.8, 1.2) {\Large \textbf{I}};

  \draw[->, thick, blue] (3, -1) -- (0.5, 1.5);
  \node[blue, right, xshift=2pt] at (3.0, -1) {$\ket{\psi}_F$};
  \node[red, left, xshift=-2pt] at (0.5, 1.5) {$V\ket{\psi}$};

  \node at (1.5, -0.6) {$\rho_E(\psi)$};
  \node at (-0.2, 0.8) {$\rho_I(\psi)$};

\end{tikzpicture}
\caption{An infalling state $\ket{\psi}$ in a fixed sector $\mathcal{H}_F^{(c)}$ crosses the horizon of a black hole prepared in $\ket{\Phi_0}$. A single unitary maps the pair to interior and exterior; the interior channel $\mathcal{N}_I$ and its complement $\mathcal{N}_E$ are two faces of the same isometry $W$. Assumption (A3) states that $\mathcal{N}_I$ is $\varepsilon_{\diamond}$-close to an isometry; the theorem bounds what $\mathcal{N}_E$ can then reveal.}
\label{fig:setup}
\end{figure}

\textit{Result.---}Our main result classifies the horizons compatible with a given amount of microstate hair.

\textbf{Theorem (horizon classification).} \emph{Under (A1)--(A3), within a fixed sector, $\varepsilon\equiv 1-\min_{\ket{\psi}}F_I(\psi)\leq\varepsilon_{\diamond}$ is the worst-case interior fidelity deficit, and the classification is parametrized by $\varepsilon$. The pairs $(D_{\max},\varepsilon)$ with $\varepsilon\leq\tfrac{1}{2}$ that a horizon can realize are exactly those satisfying} Eq.~\eqref{eq:disk}, equivalently $D_{\max}\leq
2\sqrt{\varepsilon(1-\varepsilon)}$. Moreover:
\begin{itemize}
\item[(i)] \emph{No horizon lies outside this region.}
\item[(ii)] \emph{Every point of the region is realized. The boundary is realized by dephasing channels; the interior by horizons that rotate the infalling state unitarily before embedding it, which cost smoothness while adding no hair (Supplemental Material \cite{SM}, Sec.~IV).}

\item[(iii)] \emph{For $0<\varepsilon<\tfrac{1}{2}$, the boundary is realized by, and only by, horizons that act as a which-state measurement on the extremal pair: they transmit those two states undisturbed while depositing in the exterior a record of which one fell in, of overlap $1-2\varepsilon$. Their action on the rest of the sector is unconstrained beyond the worst-case fidelity hypothesis; in the qubit case, the extremal pair exhausts a basis, and the characterization is global. At $\varepsilon=0$ the exterior channel is constant and carries no which-state record.}
\end{itemize}

Part (iii) is the physical content. Figure~\ref{fig:region} displays the classification. A horizon exhibiting microstate hair at level $D_{\max}$ must pay at least $\varepsilon\geq\tfrac{1}{2} (1-\sqrt{1-D_{\max}^{2}}\,)$ in smoothness, and below maximal hair, the horizon that pays exactly that price is a measuring device: it reads the pair it distinguishes and deposits a partial which-state record outside. Hair is cheapest when the horizon measures. Part (ii) says the rest of the region is not empty but wasteful: a horizon may disturb infalling matter unitarily, paying smoothness, and gain no hair in return, sweeping out the interior of Figure~\ref{fig:region} at fixed $D_{\max}$. 

Realization here is at the level of the channel: every point is achieved by a horizon-crossing channel satisfying the setup above and the worst-case fidelity hypothesis, and which of these arises from solutions of a given gravitational theory is a separate dynamical question.

\textit{The engine.---}The Theorem follows from a trade-off bound in quantum channel theory, which we state separately as it is of independent interest.

\textbf{Lemma.} \emph{Let $\mathcal{N}_I$ be a channel whose worst-case pure-input fidelity to an isometry $V$ satisfies $F_I(\psi)\geq 1-\varepsilon$ for all pure $\ket{\psi}$, with $\varepsilon\leq\tfrac{1}{2}$. Then the complementary channel obeys}
\begin{equation}\label{eq:bound}
\tfrac{1}{2}\bigl\lVert\rho_E(\psi)-\rho_E(\psi')\bigr\rVert_1\;\leq\; 2\sqrt{\varepsilon(1-\varepsilon)}
\end{equation}
\emph{for every pair of inputs, and the bound is exactly tight.}

\textit{Proof sketch} (complete proof in the Supplemental Material \cite{SM}, Sec.~II).---\textit{Step 0.} By linearity of $\mathcal{N}_E$, the maximum is attained on orthogonal pairs, so take $\expval{\psi|\psi'}=0$.

\textit{Step 1: pointer decomposition.} Write $W\ket{\psi}=\ket{V\psi}\otimes \ket{u_\psi}+\ket{\Xi_\psi}$ with $\ket{u_\psi}\equiv(\bra{V\psi}\otimes \mathbb{1}_E)W\ket{\psi}$, so that $(\bra{V\psi}\otimes\mathbb{1}_E) \ket{\Xi_\psi}=0$ and $\lVert u_\psi\rVert^{2}=F_I(\psi)$ exactly. Tracing out the interior, the cross terms vanish identically, leaving
\begin{equation}\label{eq:pointer}
\rho_E(\psi)\;=\;\ketbra{u_\psi}+G_\psi\,,\quad G_\psi\succeq 0\,,\quad
\mathrm{Tr}\,G_\psi=1-F_I(\psi)\,:
\end{equation}
a subnormalized pure ``pointer'' plus a positive remainder, with no interference between them.

\textit{Step 2: phase-averaged overlap bound.} Feeding the superpositions $(\ket{\psi}+e^{i\phi}\ket{\psi'})/\sqrt{2}$ into the fidelity hypothesis and averaging over $\phi$ isolates the pointer overlap; the fidelity deficits of the two states cancel exactly, and
\begin{equation}\label{eq:overlap}
\mathrm{Re}\expval{u_\psi|u_{\psi'}}\;\geq\;1-2\varepsilon\,.
\end{equation}

\textit{Step 3: assembly.} With $s\equiv[1-F_I(\psi)]+[1-F_I(\psi')]$, the exact trace norm of a difference of rank-one operators, together with Eqs.~\eqref{eq:pointer} and \eqref{eq:overlap}, gives
\begin{equation}\label{eq:assembly}
\tfrac{1}{2}\bigl\lVert\rho_E(\psi)-\rho_E(\psi')\bigr\rVert_1\;\leq\;
\tfrac{1}{2}\sqrt{(2-s)^{2}-4(1-2\varepsilon)^{2}}\;+\;\tfrac{s}{2}\,,
\end{equation}
whose right-hand side is non-increasing in $s$; its maximum at $s=0$ equals $2\sqrt{\varepsilon(1-\varepsilon)}$. \hfill$\square$

Part (i) of the Theorem is the Lemma applied to the horizon channel. Part (iii) follows from the equality analysis: for $0<\varepsilon<\tfrac{1}{2}$ the right-hand side of Eq.~\eqref{eq:assembly} is \emph{strictly} decreasing in $s$, so equality forces $s=0$ and a real pointer overlap $1-2\varepsilon$, which is precisely the which-state measurement structure; the degenerate endpoint $\varepsilon=\tfrac{1}{2}$, where the bound reduces to the trivial ceiling $D_{\max}\leq 1$, is treated in Supplemental Material \cite{SM}, Sec.~III. Part (ii) follows by composing the saturating dephasing channel at parameter $q$ with a unitary $U$ acting on the infalling system before embedding: $D_{\max}$ is unchanged, since pre-composition with a unitary merely relabels the extremal pair, while $\varepsilon$ increases continuously from $q$ as $U$ departs from the identity, realizing every point with the same $D_{\max}$ and larger $\varepsilon$ (Supplemental Material \cite{SM}, Sec.~IV).

\textit{The price of hair.---}Inverting Eq.~\eqref{eq:disk} on $\varepsilon\leq\tfrac{1}{2}$, and noting that the statement is trivial for $\varepsilon\geq\tfrac{1}{2}$, gives the smoothness cost of microstate hair, valid for all $\varepsilon$:
\begin{equation}\label{eq:tradeoff}
\boxed{\;\varepsilon\;\geq\;\tfrac{1}{2}\Bigl(1-\sqrt{1-D_{\max}^{2}}\,
\Bigr)\;}
\end{equation}

\textbf{Corollary 1 (exact threshold).} Perfect exterior discrimination of some same-sector pair, $D_{\max}=1$, requires $\varepsilon\geq\tfrac{1}{2}$, the disruption of a complete measurement on that pair; resolving the full microstate basis requires at least as much. Conversely, at the nonperturbative smoothness expected of a semiclassical horizon, $\varepsilon\sim e^{-\alpha S_{\rm BH}}$, Eq.~\eqref{eq:disk} permits only $D_{\max}\leq 2e^{-\alpha S_{\rm BH}/2}$, which for $\alpha=1$ is parametrically compatible with the $e^{-S_{\rm BH}/2}$ scale of eigenstate-thermalization off-diagonal matrix elements for simple observables \cite{deutsch1991quantum,srednicki1994chaos}, though that ansatz does not by itself determine the unrestricted quantity $D_{\max}$.

\textbf{Corollary 2 (the minimal horizon is a measurement).} For $0<D_{\max}<1$, among all horizons exhibiting that microstate distinguishability, those of least disruption act as which-state measurements on the pair they distinguish, nondemolition on that pair: they leave those two states entirely undisturbed and disturb only their superpositions. At $D_{\max}=1$, a complete measurement attains the threshold $\varepsilon=1/2$, but the minimizer is no longer unique. The least a horizon can do to have hair is to measure what falls in.

\begin{figure}[t]
\centering
\begin{tikzpicture}
\begin{axis}[
  width=0.92\columnwidth, height=0.74\columnwidth,
  xlabel={horizon disruption $\varepsilon$},
  ylabel={microstate hair $D_{\max}$},
  xmin=0, xmax=0.5, ymin=0, ymax=1.12,
  xtick={0,0.1,0.2,0.3,0.4,0.5},
  xticklabels={$0$,$0.1$,$0.2$,$0.3$,$0.4$,$1/2$},
  ytick={0,0.25,0.5,0.75,1.0},
  axis line style={line width=1.0pt},
  tick style={line width=1.0pt, black},
  major tick length=3.5pt,
  label style={font=\small}, tick label style={font=\footnotesize},
  clip=false]

\addplot[draw=none, fill=blue!10, domain=0:0.5, samples=300]
        {2*sqrt(x*(1-x))} \closedcycle;

\addplot[name path=env, draw=none, domain=0:0.5, samples=300] {2*sqrt(x*(1-x))};
\addplot[name path=top, draw=none, domain=0:0.5, samples=2] {1.12};
\addplot[fill=black!12, draw=none] fill between[of=env and top];

\addplot[black, line width=1.6pt, domain=0:0.5, samples=400] {2*sqrt(x*(1-x))};

\addplot[only marks, mark=*, mark size=2.2pt, black] coordinates {(0.5,1.0)};

\addplot[only marks, mark=*, mark size=1.8pt, black!70]
        coordinates {(0.0670,0.5)};
\draw[->, line width=0.9pt, black!70]
      (axis cs:0.0670,0.5) -- (axis cs:0.465,0.5);

\node[font=\footnotesize, anchor=west] at (axis cs:0.055,1.03) {forbidden};
\node[font=\footnotesize, anchor=west, rotate=20] at (axis cs:0.195,0.70)
      {measurement horizons};
\node[font=\footnotesize, anchor=west] at (axis cs:0.175,0.29)
      {nonminimal channels};
\end{axis}
\end{tikzpicture}
\caption{\label{fig:region}The horizon channels allowed by the assumptions. Microstate hair $D_{\max}$ and horizon disruption $\varepsilon$ are confined to the shaded region, $D_{\max}^{2}+(1-2\varepsilon)^{2}\leq 1$, and every point of it is realized. Below maximal disruption, its boundary is attained only by channels that measure the pair they distinguish, transmitting those two states undisturbed while recording outside which one fell in: these horizons pay the least smoothness for
the hair they exhibit. Interior points are realized by horizons that rotate the infalling state unitarily before embedding it, paying disruption while adding no hair; the arrow shows one such family, at fixed $D_{\max}$ and increasing $\varepsilon$, which sweeps the segment up to $\varepsilon=1/2$. Perfect discrimination of some pair, $D_{\max}=1$, is possible only at $\varepsilon\geq 1/2$ (dot), the operating point of a complete measurement; for $\varepsilon>1/2$ the classification is vacuous.}
\end{figure}

\textit{Classical labels versus quantum microstates.---}At $\varepsilon=0$, Eq.~\eqref{eq:disk} gives $D_{\max}=0$ exactly: no exterior observable distinguishes states within a fixed sector. Distinctions between sectors and between different backgrounds remain encoded in their exterior labels. When the theory is Einstein--Maxwell, the labels reduce to $(M, Q, J)$ and what is recovered, without the uniqueness theorems or the field equations, is the operational content of no-hair: an exactly smooth infall channel leaves no exterior record of which same-sector state fell in. The geometric uniqueness theorems, which classify stationary solutions of the field equations, are not reproduced. The classification thus cleanly separates two notions of hair: classical and topological distinctions are labels of the background or of the infalling sector and are not priced by the intra-sector quantity $D_{\max}$, while quantum microstate hair is intra-sector distinguishability, priced by Eq.~\eqref{eq:tradeoff}. Model-specific values of the distinguishability in Eq.~\eqref{eq:Dmax} have been computed directly in holographic settings \cite{kudler2021distinguishing}; Eq.~\eqref{eq:tradeoff} fixes, model-independently, what any such value costs.

\textit{Classical and topological hair as sector labels.---}The hairy solutions that motivate our question illustrate the sector structure rather than contradict the classification. Secondary hair, determined by the conserved charges, carries no data beyond them. Primary hair, an independent classically measurable parameter such as the non-Abelian data of colored black holes \cite{bizon1990colored,volkov1999gravitating} or a scalar charge \cite{herdeiro2015asymptotically}, is a property of the background: distinct values are distinct solutions, hence distinct channels, and enter $c$ only insofar as the infalling system itself carries the corresponding charge. The scalar-haired Kerr solutions \cite{herdeiro2014kerr} are the instructive case: they have regular horizons, carry a conserved global charge independent of $(M,J)$, and can be degenerate with Kerr in the asymptotic charges alone. That charge labels one stationary family against another; it records nothing about which state fell in. Discrete gauge hair \cite{krauss1989discrete,coleman1992quantum}, invisible to every classical probe, is superselected: the superposition step of the Lemma cannot be run across it, and it is accordingly a sector label as well. Hair of the background is carried by the fixed state $\ket{\Phi_0}$ and joins $c$ only when the infalling system carries the corresponding charge. In every case, such hair distinguishes distinct classical or topological solutions; what Eq.~\eqref{eq:tradeoff} prices is the intra-sector distinguishability of quantum microstates, which none of them supplies (Supplemental Material \cite{SM}, Sec.~VIII).

\textit{Fuzzballs.---}Microstate geometries replace the horizon with structure and are, in principle, individually distinguishable from outside \cite{mathur2005fuzzball}. Insofar as that structure imprints the infalling state on the exterior, so that same-sector infalling states achieve $D_{\max}>0$ through a common horizon-crossing channel, Eq.~\eqref{eq:tradeoff} applies directly, and perfect discrimination of some same-sector pair, $D_{\max}\to 1$, requires $\varepsilon\geq 1/2$. By Corollary~2, below that endpoint, the least such a construction can do at the horizon is a which-state measurement on the pair it distinguishes. This is consistent with the fuzzball program, which explicitly replaces the smooth horizon with structure, and makes the trade-off it accepts quantitative, optimal, and model-independent.

\textit{Soft hair.---}Whether supertranslation charges store microstate information is debated: dressing arguments indicate that soft modes decouple from the hard data \cite{mirbabayi2016dressed,bousso2017soft}, in which case soft hair is degenerate and contributes nothing to Eq.~\eqref{eq:Dmax}. The classification sharpens the debate into a dichotomy: either soft charges are sector labels or degenerate dressings, with $D_{\max}=0$ and no constraint incurred, or they resolve same-sector states \cite{hawking2016soft}, in which case Eq.~\eqref{eq:tradeoff} prices them, and the horizon cannot remain smooth.

\textit{Post-Page-time black holes.---}In the Hayden--Preskill regime, with rapidly mixing unitary dynamics and access to the early and newly emitted radiation \cite{hayden2007black}, $D_{\max}$ approaches unity and Eq.~\eqref{eq:tradeoff} forces $\varepsilon\geq 1/2$ up to exponentially small corrections, provided the bipartition continues to hold there. That proviso is precisely what the modern resolutions of the Page-curve problem deny.

\textit{Pre-existing exterior records.---}Within this framework, at exact smoothness any exterior distinguishability must already be present in the marginal of an exterior reference before infall, correlations differing only in the inaccessible purification being unrevealed: pure inputs yield $D_{\max}=0$ at $\varepsilon=0$, so any distinguishability there must reside in correlations that were external to the black hole all along, and the exterior thereby accesses only what it already held (Supplemental Material \cite{SM}, Sec.~IX).

\textit{Discussion.---}The classification holds in any semiclassical theory of gravity satisfying (A1)--(A3) within a fixed sector: higher-dimensional, string-theoretic, and multi-field black objects are covered alike, with the form and constants of the classification independent of dimension and of the gravitational dynamics, and the theory determining the sector labels of Eq.~\eqref{eq:sectors}, the ideal infall map, the realized horizon channel, and hence which point inside the allowed region is attained. It converts the complementarity principle that no observer sees both copies \cite{susskind1993stretched} into the achieved boundary of an allowed region, together with a physical identification of the horizons that sit on it.

\textit{What ``measurement'' does and does not mean.---}The parameter $\varepsilon$ is an operational measure of the departure of infall from an information-preserving isometry, and Corollary~2 is correspondingly a statement about information flow, not about energy. The extremal horizon reads the pair it distinguishes and records which one crossed; it is nondemolition on that pair, and nothing in the kinematics fixes its stress tensor. We therefore do not claim that minimal microstate hair implies an energetic curtain of the kind envisaged in firewall scenarios \cite{almheiri2013black}, a distinct and stronger statement that requires dynamical input beyond our assumptions. What we establish is the informational content of the minimal horizon: it measures.

Four further discussions are provided in the End Matter. Appendix A places the Lemma against the existing continuity and duality bounds, which it improves for the pairwise quantity considered here, and names the wave-particle-duality antecedent of the envelope. Appendix B examines the robustness of the smoothness hypothesis under restricted probe classes and under complexity-bounded notions of smoothness. Appendix C relates the classification to horizon decoherence and to the question of whether gravitational dressing imposes a kinematic floor on $\varepsilon$. Appendix D distinguishes the classification from computational censorship.

\textit{Scope.---}We consider macroscopic, semiclassical black holes with approximately sharp horizons, away from extremality; the effective bipartition is assumed on the chosen leading-order semiclassical code subspace (Supplemental Material \cite{SM}, Sec.~VII), and the result is instantaneous, relating interior fidelity to exterior distinguishability at fixed semiclassical time after infall, with the age of the black hole entering only through the post-Page-time regime discussed above. Finally, $\varepsilon$ quantifies the total departure of infall from an isometry, whatever its origin: in the vacuum-exterior idealization adopted here, horizon-scale physics is the only available source, while in astrophysical environments, ambient interactions contribute to $\varepsilon$ and to $D_{\max}$ alike, consistently with the classification.

\textit{Outlook.---}Three directions are natural: computing the dressing-induced distinguishability of same-sector states in linearized gravity, which would convert the parametric floor of Appendix C into a theorem; determining the energetics of the extremal which-state horizon, which requires dynamical input that the present kinematic analysis does not supply; and extending the analysis to the post-Page-time regime, where exterior decodability is guaranteed and the classification runs in the reverse direction.

\textit{Data availability.---}No data were created or analyzed in this study.

\nocite{witten2018notes,soni2024type,almheiri2015bulk,pastawski2015holographic,beny2010general,fuchs1999cryptographic,crann2016private,shirokov2022optimal,penington2020entanglement,almheiri2019entropy}

\bibliographystyle{apsrev4-2}
\bibliography{MBL}

\onecolumngrid
\vspace{12pt}
\begin{center}
\rule{0.45\textwidth}{0.4pt}\\[8pt]
{\bf End Matter}
\end{center}
\vspace{6pt}
\twocolumngrid

\section*{Appendix A: Relation to prior bounds}

The Lemma sharpens a known family of estimates. For the reference-free pairwise distinguishability considered here, it gives the optimal consequence of a worst-case pure-state fidelity hypothesis, improving on the bound obtained by routing the problem through the general channel-continuity theorem of Kretschmann, Schlingemann, and Werner \cite{kretschmann2008information}, whose square-root scaling cannot be improved in that generality, the constant being established for the modified form of the conjecture and proved for Kraus-rank-one channels \cite{vomende2023progress}, and the corresponding square-root bounds, in their own diamond-norm parameters, obtained from the private-correctable duality \cite{kretschmann2008complementarity} and from complementary-observable transmission \cite{renes2014operationally}. Tight trade-offs are known for a different pair of quantities, measurement error against deviation from the identity \cite{hashagen2019universality}; the envelope of Eq.~\eqref{eq:disk} is not derivable from those, and the full comparison, including the demonstration that arguments routed through the ideal constant output cannot reach the envelope, is given in the Supplemental Material \cite{SM}, Sec.~VI. 

The envelope of Eq.~\eqref{eq:disk} has an older antecedent worth naming: under the identification of $1-2\varepsilon$ with fringe visibility it coincides with the wave-particle duality relation $D^{2}+V^{2}\leq 1$ of Jaeger, Shimony, and Vaidman and of Englert \cite{jaeger1995two,englert1996fringe}, later shown equivalent to an entropic uncertainty relation \cite{coles2014equivalence}. The content of the Lemma is that a worst-case fidelity hypothesis over an entire sector, with a general isometry as the ideal and no assumed interaction structure, forces the envelope for all channels and, at equality, recovers the undisturbed two-path structure that the interferometric setting assumes at the outset. The horizon reading of the coincidence is direct: a black hole exhibiting microstate hair is a which-way detector for what falls in.

The square-root scaling at small $\varepsilon$ is physical rather than an artifact of proof technique: which-state information is carried at the amplitude level while interior disruption accrues at the probability level, so a leak of amplitude $\delta$ purchases distinguishability $D\sim\delta$ at fidelity cost $\varepsilon\sim\delta^{2}$. Computations that retain only the populations of the complementary channel miss this and suggest a spurious linear scaling; the coherences dominate (Supplemental Material \cite{SM}, Sec.~V).

\section*{Appendix B: Robustness of the smoothness hypothesis}

The parameter $\varepsilon$ is worst-case over the sector by design, since the equivalence principle is a statement about all infalling states, and the proof uses the hypothesis only on the two-dimensional span of the extremal pair (Supplemental Material \cite{SM}, Sec.~II, Remark (i)). Constructions in which horizon microstructure is invisible to restricted probe classes \cite{burman2024bottom} are therefore constrained only to the extent that their exterior distinguishability survives the worst-case optimization; smoothness of low-point correlators for typical probes is compatible with the classification.

A further weakening replaces worst-case smoothness with smoothness against computationally bounded observers: constructions of the interior as a non-isometric code protected by complexity are drastically non-isometric as maps while remaining indistinguishable from isometric infall to any sub-exponential operation \cite{akers2024black}. Our $\varepsilon$ is a property of the channel rather than of an observer's resources, and the classification constrains whether microstate hair exists given channel-level smoothness, not whether it can be detected or decoded.

\section*{Appendix C: Horizon decoherence and gravitational dressing}

A dynamical counterpart of this informational reading is due to Danielson, Satishchandran, and Wald \cite{danielson2022black,danielson2023killing}, who show that the long-range field of a body held in superposition outside a black hole is registered on the horizon, necessitating a flux of soft quanta through it, so that the horizon harvests which-path information about the superposition; the effect is reproduced by an ordinary body at finite temperature with matching low-frequency fluctuations \cite{biggs2024comparing}. Our statement runs in the opposite direction and is kinematic rather than rate-based: it concerns what the exterior learns about matter that crosses, assumes no mechanism, and is an exact envelope on a single crossing event.

A recent channel-theoretic formulation of the horizon information--decoherence correspondence was given by Danielson and Satishchandran \cite{danielson2025horizons}. They define a recovery-optimized channel decoherence and a fidelity-based measure of the complementary channel's departure from its best constant output, and prove an exact equality between them for arbitrary channels. The quantities classified here are different on both sides. Our disturbance is measured relative to a fixed infall isometry, with no recovery optimization, while our information quantity is the direct pairwise trace distance between complementary outputs. Thus, a recoverable unitary misalignment has zero decoherence in their sense but generally positive $\varepsilon$ in ours, and their equality does not directly determine the attainable $(D_{\max},\varepsilon)$ region. Indeed, the fixed-output pivot-and-triangle route of Supplemental Material \cite{SM}, Sec.~VI, is floored at $2\sqrt{\varepsilon}$ on our saturating family, strictly above the exact envelope $2\sqrt{\varepsilon(1-\varepsilon)}$. Their finite-time horizon estimate shares the functional form $\tfrac{1}{2}(1-\sqrt{1-\,\cdot\,}\,)$ with the inversion of Eq.~(1), both descending from fidelity geometry, but is explicitly non-tight and comes with no saturating family. The present result supplies the exact attainable region for the fixed-isometry and pairwise-distinguishability quantities, together with its saturating family and nondegenerate equality class.

Whether $\varepsilon=0$ is attainable in gravity connects the classification to a live debate. If gravitational constraints make information available near the boundary \cite{laddha2021holographic,raju2022failure}, same-sector states are kinematically distinguishable through their dressing at the level $D\sim E/M_{\rm Pl}$ for an infalling system of energy $E$, and Eq.~\eqref{eq:tradeoff} implies a floor $\varepsilon\gtrsim(E/M_{\rm Pl})^{2}/4$, of order $1/S_{\rm BH}$ for Hawking-scale quanta; if information can instead be localized by dressing to features of the state \cite{donnelly2018gravitational,bahiru2024holography}, no kinematic floor arises. Within our assumptions, exact smoothness requires the absence of gravitationally accessible same-sector information; localizability removes that kinematic obstruction without by itself guaranteeing $\varepsilon=0$.

\section*{Appendix D: Relation to computational censorship}

The Hayden--Preskill mechanism shows that information thrown into an old black hole becomes encoded in the radiation \cite{hayden2007black}; the decoding tasks relevant to firewall arguments can require exponential computational resources \cite{harlow2013quantum}. Our result is logically independent: at $\varepsilon=0$ the classification forbids \emph{any} exterior protocol, efficient or not, from achieving $D_{\max}>0$, and for $\varepsilon\ll 1$ it bounds every such protocol by $D_{\max}\leq 2\sqrt{\varepsilon(1-\varepsilon)}$. Computational censorship concerns the complexity of decoding hair that is in principle present; Eq.~\eqref{eq:tradeoff} concerns whether such hair exists at all given smoothness, and neither statement implies the other.

\clearpage
\pagebreak
\toggletrue{issupplement}  

\onecolumngrid
\renewcommand{\thesection}{S\arabic{section}}
\renewcommand{\theequation}{S\arabic{equation}}
\setcounter{section}{0}
\setcounter{equation}{0}

\vspace*{1cm}
\begin{center}
  {\LARGE \bf Supplemental Material}\\[2.5em]
  {\large \bf Black-Hole Microstate Hair Requires a Horizon Measurement}\\[1.5em]
  {\normalsize Sudhanva Joshi\textsuperscript{*1} and Sunil Kumar Mishra\textsuperscript{‡1}}\\[1em]
  {\footnotesize
  \textsuperscript{1} Department of Physics, Indian Institute of Technology (Banaras Hindu University), Varanasi - 221005, India \\
  }
\end{center}
\vspace{1cm}

\maketitle
This Supplemental Material provides all constructions and proofs deferred in the main text. Equations of the form (1)--(8) refer to the main text; equations here are labeled (S1), (S2), and so forth. For navigation: the hypothesis hierarchy, showing that only the worst-case pure-input fidelity is used and that the diamond-norm assumption (A3) implies it, is given in Sec.~\ref{sec:hierarchy}; the complete proof of the Lemma, together with its use only on the two-dimensional span of the extremal pair and its independence of all dimensions and of the choice of dilation, in Sec.~\ref{sec:proof}; the equality analysis establishing part (iii) of the Theorem, the which-state saturating family, and the degenerate endpoint, in Sec.~\ref{sec:equality}; the construction realizing the interior of the allowed region, part (ii) of the Theorem, in Sec.~\ref{sec:interior}; the demonstration that the coherences of the complementary channel dominate its populations, in Sec.~\ref{sec:examples}; the comparison with prior continuity, duality, and tight-trade-off bounds, including the proof that arguments routed through the ideal constant output cannot reach the envelope, in Sec.~\ref{sec:prior}; the effective interior/exterior factorization, its field-theoretic basis, and the additional assumptions required in gravity, in Sec.~\ref{sec:factorization}; the sector structure and the necessity of the sector restriction, in Sec.~\ref{sec:sectors}; and the analysis of pre-existing entanglement, in Sec.~\ref{sec:entanglement}.

\vspace{6pt}
\noindent\textbf{Lemma} (main text). \emph{Let $\mathcal{N}_I$ have worst-case pure-input fidelity $F_I(\psi)\geq 1-\varepsilon$ to an isometry $V$, with $\varepsilon\leq 1/2$. Then $\tfrac{1}{2}\lVert\rho_E(\psi)-\rho_E(\psi')\rVert_1\leq 2\sqrt{\varepsilon(1-\varepsilon)}$ for every pair, and the bound is exactly tight.}

\section{The hypothesis hierarchy}
\label{sec:hierarchy}

The interior channel and its complement are defined by the Stinespring isometry $W:\mathcal{H}_F^{(c)}\to\mathcal{H}_I\otimes\mathcal{H}_E$, $W\ket{\psi}=U(\ket{\psi}\otimes\ket{\Phi_0})$, via $\mathcal{N}_I(\rho)=\mathrm{Tr}_E[W\rho W^{\dagger}]$ and $\mathcal{N}_E(\rho)=\mathrm{Tr}_I[W\rho W^{\dagger}]$, with $\rho_E(\psi)\equiv\mathcal{N}_E(\ketbra{\psi})$. Throughout, $\mathcal{H}_F$ denotes the fixed sector $\mathcal{H}_F^{(c)}$.

Assumption (A3) of the main text is stated in diamond norm. Evaluating it on the pure input $\ketbra{\psi}$ and using the variational characterization of the trace distance with the test operator $P=V\ketbra{\psi}V^{\dagger}$,
\begin{equation}\label{eq:S_hierarchy}
1-F_I(\psi)\;=\;\mathrm{Tr}\bigl[P\,(V\ketbra{\psi}V^{\dagger}-\mathcal{N}_I(\ketbra{\psi}))\bigr]\;\leq\;\tfrac{1}{2}\bigl\lVert \mathcal{N}_I(\ketbra{\psi})-V\ketbra{\psi}V^{\dagger}\bigr\rVert_1\;\leq\;\varepsilon_{\diamond}\,,
\end{equation}
so (A3) implies $F_I(\psi)\geq 1-\varepsilon_{\diamond}$ for every pure $\ket{\psi}$, that is $\varepsilon\leq\varepsilon_{\diamond}$ for the worst-case fidelity deficit $\varepsilon$. The proofs below use \emph{only} this state-level consequence; the diamond-norm form is adopted in the main text as the physically motivated statement of the equivalence principle, robust to entanglement-assisted probes. The converse implication carries dimensional factors and is nowhere needed. For $\varepsilon_{\diamond}\leq\tfrac{1}{2}$, monotonicity of $2\sqrt{x(1-x)}$ on $[0,\tfrac{1}{2}]$ gives the corresponding upper envelope with $\varepsilon_{\diamond}$ in place of $\varepsilon$. Independently of the size of $\varepsilon_{\diamond}$, every lower bound derived for $\varepsilon$ applies a fortiori to $\varepsilon_{\diamond}$, since $\varepsilon\leq\varepsilon_{\diamond}$. The saturating family of Sec.~\ref{sec:equality} satisfies both readings with the same value.

We record the identity on which the proof is built. Expanding the partial trace in an orthonormal basis $\{\ket{e_k}\}$ of $\mathcal{H}_E$,
\begin{equation}\label{eq:S_identity}
F_I(\psi)\;=\;\bra{V\psi}\,\mathcal{N}_I(\ketbra{\psi})\,\ket{V\psi}\;=\;\sum_k\bigl|(\bra{V\psi}\otimes\bra{e_k})\,W\ket{\psi}\bigr|^{2}\;=\;\bigl\lVert(\bra{V\psi}\otimes\mathbb{1}_E)\,W\ket{\psi}\bigr\rVert^{2}\,.
\end{equation}

\section{Proof of the Lemma}
\label{sec:proof}

\subsection{Step 0: reduction to orthogonal pairs}

Let $\ket{\psi},\ket{\psi'}$ be pure with $t=\ip{\psi}{\psi'}$. The operator $\ketbra{\psi}-\ketbra{\psi'}$ is Hermitian, traceless, of rank at most two, with $\mathrm{Tr}[(\ketbra{\psi}-\ketbra{\psi'})^{2}]=2(1-|t|^{2})$; its eigenvalues are therefore $\pm\sqrt{1-|t|^{2}}$, and
\begin{equation}\label{eq:S_reduction}
\ketbra{\psi}-\ketbra{\psi'}\;=\;\sqrt{1-|t|^{2}}\,\bigl(\ketbra{P_+}-\ketbra{P_-}\bigr)
\end{equation}
for orthonormal pure $\ket{P_\pm}\in\mathrm{span}\{\ket{\psi},\ket{\psi'}\}\subseteq\mathcal{H}_F^{(c)}$. By linearity of $\mathcal{N}_E$, the pair distance equals $\sqrt{1-|t|^{2}}$ times that of the orthogonal pair, so the maximum defining $D_{\max}$ in Eq.~(3) of the main text is attained on an orthogonal pair. We take $\ip{\psi}{\psi'}=0$ and write $F\equiv F_I(\psi)$, $F'\equiv F_I(\psi')$, $\delta\equiv 1-F$, $\delta'\equiv 1-F'$.

\subsection{Step 1: pointer decomposition}

Define
\begin{equation}\label{eq:S_pointer_def}
\ket{u_\psi}\;\equiv\;(\bra{V\psi}\otimes\mathbb{1}_E)\,W\ket{\psi}\;\in\;\mathcal{H}_E\,,\qquad
\ket{\Xi_\psi}\;\equiv\;W\ket{\psi}-\ket{V\psi}\otimes\ket{u_\psi}\,.
\end{equation}
By construction $(\bra{V\psi}\otimes\mathbb{1}_E)\ket{\Xi_\psi}=0$, so the two pieces are orthogonal, and by Eq.~\eqref{eq:S_identity}, $\lVert u_\psi\rVert^{2}=F$ and $\lVert\Xi_\psi\rVert^{2}=\delta$. Tracing out the interior, the cross terms vanish identically: for any $\ket{e_a},\ket{e_b}$,
\begin{equation}
\bra{e_a}\,\mathrm{Tr}_I\bigl[(\ket{V\psi}\otimes\ket{u_\psi})\bra{\Xi_\psi}\bigr]\,\ket{e_b}\;=\;\ip{e_a}{u_\psi}\;\overline{(\bra{V\psi}\otimes\bra{e_b})\ket{\Xi_\psi}}\;=\;0\,.
\end{equation}
Hence
\begin{equation}\label{eq:S_pointer}
\rho_E(\psi)\;=\;\ketbra{u_\psi}\;+\;G_\psi\,,\qquad G_\psi\;\equiv\;\mathrm{Tr}_I\bigl[\ketbra{\Xi_\psi}\bigr]\;\succeq\;0\,,\qquad\mathrm{Tr}\,G_\psi=\delta\,,
\end{equation}
which is Eq.~(5) of the main text.

\subsection{Step 2: the phase-averaged overlap bound}

For $\phi\in[0,2\pi)$ let $\ket{c_\phi}=(\ket{\psi}+e^{i\phi}\ket{\psi'})/\sqrt{2}$. Using linearity of $W$ and $V$ and $\ip{V\psi}{V\psi'}=0$,
\begin{equation}\label{eq:S_AB}
(\bra{Vc_\phi}\otimes\mathbb{1}_E)\,W\ket{c_\phi}\;=\;\underbrace{\tfrac{1}{2}\bigl(\ket{u_\psi}+\ket{u_{\psi'}}\bigr)}_{\ket{A}}\;+\;\underbrace{\tfrac{1}{2}\bigl(e^{i\phi}\ket{w'}+e^{-i\phi}\ket{w}\bigr)}_{\ket{B_\phi}}\,,
\end{equation}
with leaks $\ket{w}\equiv(\bra{V\psi'}\otimes\mathbb{1}_E)\ket{\Xi_\psi}$, $\ket{w'}\equiv(\bra{V\psi}\otimes\mathbb{1}_E)\ket{\Xi_{\psi'}}$ obeying $\lVert w\rVert^{2}\leq\delta$, $\lVert w'\rVert^{2}\leq\delta'$, since projections contract norms. By Eq.~\eqref{eq:S_identity} the hypothesis applied to $\ket{c_\phi}$ reads $\lVert A+B_\phi\rVert^{2}\geq 1-\varepsilon$ for every $\phi$. Averaging over $\phi$, the cross term $\ip{A}{B_\phi}$ carries $e^{\pm i\phi}$ and the interference inside $\lVert B_\phi\rVert^{2}$ carries $e^{\mp 2i\phi}$; both integrate to zero, leaving $\lVert A\rVert^{2}+\tfrac{1}{4}(\lVert w\rVert^{2}+\lVert w'\rVert^{2})\geq 1-\varepsilon$. Substituting $\lVert A\rVert^{2}=\tfrac{1}{4}[F+F'+2\,\mathrm{Re}\ip{u_\psi}{u_{\psi'}}]$ and the leak bounds,
\begin{equation}
\tfrac{1}{4}\bigl[(1-\delta)+(1-\delta')+2\,\mathrm{Re}\ip{u_\psi}{u_{\psi'}}\bigr]\;+\;\tfrac{1}{4}\bigl(\delta+\delta'\bigr)\;\geq\;1-\varepsilon\,,
\end{equation}
in which the fidelity deficits cancel \emph{exactly}, leaving
\begin{equation}\label{eq:S_overlap}
\mathrm{Re}\ip{u_\psi}{u_{\psi'}}\;\geq\;1-2\varepsilon\;\geq\;0\,,
\end{equation}
which is Eq.~(6) of the main text; the final inequality uses $\varepsilon\leq\tfrac{1}{2}$.

\subsection{Step 3: assembly}

For $P=\ketbra{u_\psi}$, $Q=\ketbra{u_{\psi'}}$, the difference $M=P-Q$ is supported on $\mathrm{span}\{u_\psi,u_{\psi'}\}$ with $\mathrm{Tr}\,M=F-F'$ and $\lambda_+\lambda_-=|\ip{u_\psi}{u_{\psi'}}|^{2}-FF'\leq 0$ by Cauchy--Schwarz, so the eigenvalues have opposite signs and
\begin{equation}\label{eq:S_ranknorm}
\lVert M\rVert_1\;=\;\lambda_+-\lambda_-\;=\;\sqrt{(F+F')^{2}-4\,|\ip{u_\psi}{u_{\psi'}}|^{2}}\,.
\end{equation}
Write $s\equiv\delta+\delta'\in[0,2\varepsilon]$, so $F+F'=2-s$. Using Eq.~\eqref{eq:S_pointer} for both states, $\lVert G_\psi-G_{\psi'}\rVert_1\leq\mathrm{Tr}\,G_\psi+\mathrm{Tr}\,G_{\psi'}=s$, and $|\ip{u_\psi}{u_{\psi'}}|\geq\mathrm{Re}\ip{u_\psi}{u_{\psi'}}\geq 1-2\varepsilon$,
\begin{equation}\label{eq:S_assembly}
\tfrac{1}{2}\bigl\lVert\rho_E(\psi)-\rho_E(\psi')\bigr\rVert_1\;\leq\;\tfrac{1}{2}\lVert M\rVert_1+\tfrac{s}{2}\;\leq\;f(s)\;\equiv\;\tfrac{1}{2}\sqrt{(2-s)^{2}-4(1-2\varepsilon)^{2}}\;+\;\tfrac{s}{2}\,,
\end{equation}
which is Eq.~(7) of the main text; the radicand is nonnegative since $s\leq 2\varepsilon$ gives $2-s\geq 2-2\varepsilon\geq 2(1-2\varepsilon)$. Because $\sqrt{x^{2}-a^{2}}\leq x$ for $x>0$,
\begin{equation}\label{eq:S_monotone}
f'(s)\;=\;\tfrac{1}{2}\Bigl[\,1-\frac{2-s}{\sqrt{(2-s)^{2}-4(1-2\varepsilon)^{2}}}\,\Bigr]\;\leq\;0\,,
\end{equation}
so $f$ is non-increasing with maximum at $s=0$:
\begin{equation}
\tfrac{1}{2}\bigl\lVert\rho_E(\psi)-\rho_E(\psi')\bigr\rVert_1\;\leq\;f(0)\;=\;\sqrt{1-(1-2\varepsilon)^{2}}\;=\;2\sqrt{\varepsilon(1-\varepsilon)}\,,
\end{equation}
which is Eq.~(4) of the main text. Maximizing over pairs and squaring gives the region of Eq.~(1), and inverting on $[0,\tfrac{1}{2}]$, where $g(\varepsilon)=2\sqrt{\varepsilon(1-\varepsilon)}$ is a strictly increasing bijection onto $[0,1]$, gives $\varepsilon\geq\tfrac{1}{2}(1-\sqrt{1-D_{\max}^{2}})$; for $\varepsilon>\tfrac{1}{2}$ the right-hand side never exceeds $\tfrac{1}{2}$, so Eq.~(8) of the main text holds for all $\varepsilon$, with expansion $\varepsilon\geq D_{\max}^{2}/4+D_{\max}^{4}/16+\cdots$. This proves part~(i) of the Theorem. \hfill$\square$

\subsection{Remarks}
\label{sec:remarks}

\emph{(i) Locality of the hypothesis.} The hypothesis was used only on $\mathrm{span}\{\ket{\psi},\ket{\psi'}\}$: the pair and the superpositions $\ket{c_\phi}$. The Lemma therefore holds with $\varepsilon$ replaced by the worst-case fidelity deficit on the two-dimensional span of the extremal pair, the strictly stronger statement invoked in the Discussion of the main text.

\emph{(ii) Dimension independence.} No step refers to the dimensions of $\mathcal{H}_F^{(c)}$, $\mathcal{H}_I$, or $\mathcal{H}_E$, and no constant depends on them; the argument is valid on separable Hilbert spaces for any channel admitting a Stinespring dilation, with the maximum in Eq.~(3) read as a supremum.

\emph{(iii) Dilation independence.} Any two Stinespring dilations of $\mathcal{N}_I$ are related by an isometry on the environment, under which all trace distances between complementary outputs are invariant; $D_{\max}$ is therefore well defined.

\emph{(iv) The exact case.} At $\varepsilon=0$, Eq.~\eqref{eq:S_pointer} gives $\rho_E(\psi)=\ketbra{u_\psi}$ with $\lVert u_\psi\rVert=1$, and Eq.~\eqref{eq:S_overlap} gives $\ip{u_\psi}{u_{\psi'}}=1$ for every orthogonal pair, so all pointers coincide: a perfectly smooth horizon has $D_{\max}=0$ exactly.

\section{Equality analysis: the boundary is the measurement horizons}
\label{sec:equality}

This section proves part~(iii) of the Theorem.

\subsection{Equality conditions}

Fix $\varepsilon\in(0,\tfrac{1}{2})$. Since $f$ in Eq.~\eqref{eq:S_assembly} is \emph{strictly} decreasing for $\varepsilon<\tfrac{1}{2}$, equality forces $s=0$, that is $F_I(\psi)=F_I(\psi')=1$ and $G_\psi=G_{\psi'}=0$, together with $\ip{u_\psi}{u_{\psi'}}=1-2\varepsilon$ real. On the extremal pair's span, the Stinespring isometry therefore acts as
\begin{equation}\label{eq:S_extremal}
W\bigl(\alpha\ket{\psi}+\beta\ket{\psi'}\bigr)\;=\;\alpha\,\ket{V\psi}\otimes\ket{u_\psi}\;+\;\beta\,\ket{V\psi'}\otimes\ket{u_{\psi'}}\,,\qquad \ip{u_\psi}{u_{\psi'}}=1-2\varepsilon\,:
\end{equation}
The two basis states cross the horizon unharmed, while a which-state record of overlap $1-2\varepsilon$ is deposited in the exterior. Conversely, any channel of the form \eqref{eq:S_extremal} satisfies the fidelity hypothesis on every superposition and saturates it on balanced ones, since for $\ket{c}=\alpha\ket{\psi}+\beta\ket{\psi'}$,
\begin{equation}\label{eq:S_basisselective}
F_I(c)\;=\;\bigl\lVert |\alpha|^{2}\ket{u_\psi}+|\beta|^{2}\ket{u_{\psi'}}\bigr\rVert^{2}\;=\;1-4|\alpha|^{2}|\beta|^{2}\varepsilon\,,
\end{equation}
equal to one on the recorded pair and minimized at $1-\varepsilon$ on balanced superpositions; its pair distinguishability is exactly $\sqrt{1-(1-2\varepsilon)^{2}}=2\sqrt{\varepsilon(1-\varepsilon)}$. 
The boundary of the region is therefore realized by, and only by, channels acting as a which-state measurement on the extremal pair, their action elsewhere constrained only by the worst-case fidelity hypothesis, which is part~(iii); for $d=2$ the extremal pair exhausts a basis, and the characterization is global.

\subsection{The dephasing family}

The canonical realization is the qubit dephasing channel, $K_0=\sqrt{1-q}\,\mathbb{1}$, $K_1=\sqrt{q}\,Z$, $q\in[0,\tfrac{1}{2}]$, with $V=\mathbb{1}$. Its worst-case fidelity is
\begin{equation}
F_I(\psi)\;=\;\sum_k\bigl|\bra{\psi}K_k\ket{\psi}\bigr|^{2}\;=\;(1-q)+q\,\langle Z\rangle_\psi^{2}\,,
\end{equation}
minimized on the equator at $1-q$, so $\varepsilon=q$. The diamond-norm distance coincides: the Choi operators satisfy $J(\mathcal{N})-J(\mathrm{id})=q(\ketbra{\Omega_Z}-\ketbra{\Omega})$ with orthogonal maximally entangled $\ket{\Omega}$, $\ket{\Omega_Z}=(Z\otimes\mathbb{1})\ket{\Omega}$, giving $\tfrac{1}{2}\lVert\mathcal{N}-\mathrm{id}\rVert_{\diamond}\geq q$, while convexity, $\mathcal{N}=(1-q)\,\mathrm{id}+q\,\mathrm{Ad}_Z$, gives the matching upper bound. This coincidence is what makes the boundary claim of the Theorem parametrization-independent: the saturating family realizes the same point of the region whether $\varepsilon$ is read as the fidelity deficit or as the diamond-norm parameter. The Stinespring vectors are
\begin{equation}\label{eq:S_dephasing}
W\ket{0}=\ket{0}\otimes\ket{\chi_0}\,,\qquad W\ket{1}=\ket{1}\otimes\ket{\chi_1}\,,\qquad \ket{\chi_{0,1}}=\sqrt{1-q}\,\ket{e_0}\pm\sqrt{q}\,\ket{e_1}\,,
\end{equation}
of the extremal form \eqref{eq:S_extremal} with $\ip{\chi_0}{\chi_1}=1-2q$, and
\begin{equation}
\tfrac{1}{2}\bigl\lVert\rho_E(0)-\rho_E(1)\bigr\rVert_1\;=\;\sqrt{1-(1-2q)^{2}}\;=\;2\sqrt{q(1-q)}\;=\;2\sqrt{\varepsilon(1-\varepsilon)}\,,
\end{equation}
saturating at every $\varepsilon\in[0,\tfrac{1}{2}]$. At $q=\tfrac{1}{2}$ the record is perfect, $\ip{\chi_0}{\chi_1}=0$, and $D_{\max}=1$.

The $d$-level generalization $K_0=\sqrt{1-(d-1)q}\,\mathbb{1}$, $K_k=\sqrt{q}\,Z^{k}$ ($k=1,\dots,d-1$, $Z$ the clock operator, $q\leq 1/(d-1)$) has worst-case fidelity attained on the uniform superposition, giving $\varepsilon=(d-1)q$, and computational-basis pairs with pure exterior outputs of pairwise overlap $1-dq$ (using $\sum_{k=1}^{d-1}\omega^{(l-j)k}=-1$ for $l\neq j$), so $D=\sqrt{dq(2-dq)}$. Comparing with the envelope at small $q$, the ratio is $\sqrt{d/(2(d-1))}$, equal to one only for $d=2$: within this family, the qubit member is extremal and higher $d$ strictly under-saturate.

Saturation is nonetheless available in every dimension. Let $Z$ be any Hermitian unitary on $\mathcal{H}_F^{(c)}$ with both eigenvalues $\pm 1$ present, and set $\mathcal{N}_q(\rho)=(1-q)\rho+qZ\rho Z$. Its worst-case fidelity is $F_I(\psi)=(1-q)+q\langle Z\rangle_\psi^{2}$, minimized at $1-q$ on any balanced superposition of a $+1$ and a $-1$ eigenstate, so $\varepsilon=q$. Taking the extremal pair to be one eigenstate from each eigenspace, both cross undisturbed, and their exterior records are $\sqrt{1-q}\,\ket{e_0}\pm\sqrt{q}\,\ket{e_1}$, of overlap $1-2q$, whence $D_{\max}=2\sqrt{q(1-q)}=2\sqrt{\varepsilon(1-\varepsilon)}$. The boundary is therefore attained for every $d\geq 2$, the qubit family above being the case $d=2$; a unitary acting on the two-dimensional extremal subspace then supplies the interior construction of Sec.~\ref{sec:interior} in any dimension.

\subsection{The degenerate endpoint}

At $\varepsilon=\tfrac{1}{2}$, $f(s)=\tfrac{1}{2}(2-s)+\tfrac{s}{2}=1$ identically, the bound reduces to the trivial ceiling $D_{\max}\leq 1$, and it is saturated by any channel whose complementary channel perfectly distinguishes some orthogonal pair, not only by measurements. An explicit witness is the completely depolarizing qubit channel $\mathcal{N}(\rho)=\tfrac{1}{2}\mathbb{1}\,\mathrm{Tr}\rho$, with Kraus operators $\{\tfrac{1}{2}\sigma_k\}$: its fidelity is $\tfrac{1}{2}$ for every pure state, so $\varepsilon=\tfrac{1}{2}$ with $s=1\neq 0$, while its complementary channel $[\widetilde{\mathcal{N}}(\rho)]_{kl}=\tfrac{1}{4}\mathrm{Tr}[\sigma_k\rho\sigma_l]$ distinguishes $\ket{0}$ from $\ket{1}$ perfectly, the difference of outputs having off-diagonal entries of magnitude $\tfrac{1}{2}$ in both the $(0,3)$ and $(1,2)$ blocks and hence trace distance one. This is why part~(iii) of the Theorem is scoped to $\varepsilon<\tfrac{1}{2}$.

\section{The interior of the allowed region}
\label{sec:interior}
This section proves part~(ii) of the Theorem: every point of the region of Eq.~(1) is realized, the boundary by the family of Sec.~\ref{sec:equality} and the interior by horizons that rotate the infalling state unitarily before embedding it.
Throughout this section, the parameter $\varepsilon$ is the worst-case interior fidelity deficit, $\varepsilon=1-\min_{\ket{\psi}}F_I(\psi)$, which is the parameter of the classification; the realization below is in that parametrization. The boundary family of Sec.~\ref{sec:equality} satisfies both the fidelity and diamond-norm readings with the same value, as shown there, so the boundary claim is parametrization-independent; the interior claim is not, and the rotated channels $\mathcal{N}_{q}\circ\mathrm{Ad}_U$ will in general have a strictly larger diamond-norm deficit than fidelity deficit.

Fix a target point $(D_0,\varepsilon_1)$ strictly inside the region, so $0\leq D_0<1$ and $\varepsilon_0<\varepsilon_1\leq\tfrac{1}{2}$, where $\varepsilon_0=\tfrac{1}{2}(1-\sqrt{1-D_0^{2}})$ is the boundary value at $D_0$. Let $\mathcal{N}_q(\rho)=(1-q)\rho+qZ\rho Z$ denote the arbitrary-dimension saturating channel of Sec.~\ref{sec:equality} at $q=\varepsilon_0$, which realizes $(D_0,\varepsilon_0)$, and consider the modified horizon
\begin{equation}\label{eq:S_rotated}
\mathcal{N}_U\;\equiv\;\mathcal{N}_{q}\circ\mathrm{Ad}_U\,,\qquad U\in SU(d)\,,
\end{equation}
which applies a unitary to the infalling system before the dephasing interaction. Two facts establish the claim.

\emph{The distinguishability is unchanged.} The complementary channel of $\mathcal{N}_U$ is $\widetilde{\mathcal{N}}_U=\widetilde{\mathcal{N}}_{(q)}\circ\mathrm{Ad}_U$, so its outputs on the pair $(\ket{\psi},\ket{\psi'})$ coincide with those of $\widetilde{\mathcal{N}}_{(q)}$ on $(U\ket{\psi},U\ket{\psi'})$. Since $U$ is a bijection of the set of pure states preserving orthogonality, maximizing over pairs gives
\begin{equation}
D_{\max}(\mathcal{N}_U)\;=\;D_{\max}(\mathcal{N}_{(q)})\;=\;D_0\,,
\end{equation}
independently of $U$: pre-composition with a unitary merely relabels which pair is extremal.

\emph{The smoothness degrades continuously.} We work in coordinates where $V=\mathbb{1}$, which is, without loss of generality, post-composition with a fixed isometry changes neither $F_I$ nor the complementary channel. The fidelity of $\mathcal{N}_U$ to the ideal is then $F_I^{U}(\psi)=\bra{\psi}\mathcal{N}_q(\ketbra{U\psi})\ket{\psi}$, whose worst case over $\psi$ defines $\varepsilon(U)$; at $U=\mathbb{1}$ this is $\varepsilon_0$. The map $U\mapsto\varepsilon(U)$ is continuous, as it is the minimum of a jointly continuous function over a compact set. Let $U_{\rm sw}\in SU(d)$ be a phase-adjusted swap, acting as $iX$ on $\mathrm{span}\{\ket{z_+},\ket{z_-}\}$ for a $+1$ eigenstate $\ket{z_+}$ of $Z$ and a $-1$ eigenstate $\ket{z_-}$, and as the identity on the orthogonal complement, so that it exchanges the two up to phases. Since $\mathcal{N}_q$ leaves every eigenstate of $Z$ invariant, $F_I^{U_{\rm sw}}(z_+)=|\ip{z_+}{z_-}|^{2}=0$, so $\varepsilon(U_{\rm sw})=1$. By the intermediate value theorem along any continuous path in $SU(d)$ from $\mathbb{1}$ to $U_{\rm sw}$, every value in $[\varepsilon_0,1]$ is attained, in particular every $\varepsilon_1\in[\varepsilon_0,\tfrac{1}{2}]$, realizing $(D_0,\varepsilon_1)$ exactly.

Since $D_0$ was arbitrary in $[0,1)$, every interior point of the region is realized; the boundary, including the endpoint $D_{\max}=1$, is supplied by the construction of Sec.~\ref{sec:equality}. Together, these give part~(ii). The physical reading is stated in the main text: horizons on the boundary measure, paying the least smoothness for the hair they exhibit, while horizons in the interior rotate unitarily, paying smoothness and gaining no hair in return.

\section{Coherences dominate populations}
\label{sec:examples}

Two standard families illustrate that the square-root scaling of the Lemma is generic and is carried by the \emph{coherences} of the complementary channel, $[\widetilde{\mathcal{N}}(\rho)]_{kl}=\mathrm{Tr}[K_k\rho K_l^{\dagger}]$ with $k\neq l$, which are of amplitude order; computations retaining only the populations miss the dominant contribution and suggest a spurious linear scaling, as stated in the main text.

\emph{(a) Depolarizing channel.} $K_0=\sqrt{1-3p/4}\,\mathbb{1}$, $K_i=\sqrt{p/4}\,\sigma_i$. Here $F_I(\psi)=(1-3p/4)+(p/4)\sum_i\langle\sigma_i\rangle_\psi^{2}=1-p/2$ for every pure state, so $\varepsilon=p/2$; the diamond-norm distance is larger, $3p/4$. For the pair $\ket{0},\ket{1}$, the populations of the complementary outputs are identical, and the difference is carried entirely by the coherences, giving
\begin{equation}
D\;=\;\sqrt{p\,(1-3p/4)}\;+\;\tfrac{p}{2}\;=\;\sqrt{2\varepsilon-3\varepsilon^{2}}\;+\;\varepsilon\,,
\end{equation}
which by Step~0 and unitary covariance equals $D_{\max}$. This lies strictly inside the region for $\varepsilon<\tfrac{1}{2}$ and meets the boundary at $\varepsilon=\tfrac{1}{2}$ ($p=1$), the degenerate point of Sec.~\ref{sec:equality}. The small-$\varepsilon$ scaling is $D_{\max}\simeq\sqrt{2\varepsilon}$, invisible to a populations-only computation, which would report only the $\mathcal{O}(\varepsilon)$ term.

\emph{(b) Amplitude-damping channel.} $K_0=\mathrm{diag}(1,\sqrt{1-\gamma})$, $K_1=\sqrt{\gamma}\,\ket{0}\!\bra{1}$. The worst-case fidelity is attained at $\ket{1}$, giving $\varepsilon=\gamma$. For the computational pair the complementary outputs are diagonal and $D(\ket{0},\ket{1})=\gamma$, linear; but for the coherence pair the complementary coherences are $\mathrm{Tr}[K_0\ketbra{\pm}K_1^{\dagger}]=\pm\sqrt{\gamma}/2$, so
\begin{equation}
D(\ket{+},\ket{-})\;=\;\sqrt{\gamma}\;=\;\sqrt{\varepsilon}\,,
\end{equation}
exactly. The linear value for the computational pair is an artifact of that pair's vanishing coherences.

\section{Relation to prior bounds}
\label{sec:prior}

\subsection{The KSW continuity theorem and the general route}
\label{sec:ksw}

The continuity theorem of Kretschmann, Schlingemann, and Werner \cite{kretschmann2008information} states that for channels $\mathcal{T}_1,\mathcal{T}_2$ with dilations into a common environment after padding the smaller environment, there is a unitary $U_{\!E}$ on the common environment with
\begin{equation}\label{eq:S_KSW}
\bigl\lVert\widetilde{\mathcal{T}}_1-\mathrm{Ad}_{U_{\!E}}\circ\widetilde{\mathcal{T}}_2\bigr\rVert_{\diamond}\;\leq\;2\sqrt{\bigl\lVert\mathcal{T}_1-\mathcal{T}_2\bigr\rVert_{\diamond}}\,,
\end{equation}
the square-root scaling being optimal in that generality \cite{kretschmann2008information,vomende2023progress}. Applied to $(\mathcal{N}_I, V(\cdot)V^{\dagger})$ with $\lVert\mathcal{N}_I-V(\cdot)V^{\dagger}\rVert_{\diamond}\leq 2\varepsilon_{\diamond}$, and noting that the complement of an isometry is the constant map $\rho\mapsto\ketbra{\sigma}\mathrm{Tr}\rho$, every exterior output lies within $\sqrt{2\varepsilon_{\diamond}}$ of the fixed operator $\widetilde{\sigma}=U_{\!E}\ketbra{\sigma}U_{\!E}^{\dagger}$, and the triangle inequality through that pivot gives
\begin{equation}\label{eq:S_KSWroute}
D_{\max}\;\leq\;2\sqrt{2\varepsilon_{\diamond}}\,.
\end{equation}
The same continuity theorem yields a related bound in the language of private and correctable subsystems: Kretschmann, Kribs, and Spekkens \cite{kretschmann2008complementarity} prove that a subsystem $\varepsilon$-correctable for a channel is $2\sqrt{\varepsilon}$-private for its complement, and conversely, where their $\varepsilon$ is an unhalved diamond-norm distance. Their argument runs through the ideal product channel exactly as above: the hypothesis is converted, by the same continuity theorem, into closeness of the complementary channel to a constant (in their terminology, deletion) map. In our normalization $\lVert\mathcal{N}_I-V(\cdot)V^{\dagger}\rVert_{\diamond}\leq 2\varepsilon_{\diamond}$, their result therefore reproduces $D_{\max}\leq 2\sqrt{2\varepsilon_{\diamond}}$; the difference in stated constants is a normalization convention, not a sharpening. Their bound is thus an instance of the barrier of Sec.~\ref{sec:pivot} at the level of proof, and like every route anchored to the ideal output, it is floored at $2\sqrt{\varepsilon}$ on the saturating family, strictly above the envelope $2\sqrt{\varepsilon(1-\varepsilon)}$. It is moreover an implication in each direction rather than a tight characterization: no optimality is claimed, no saturating family exhibited, and no equality case identified.

\subsection{Why routes through the ideal output cannot reach the envelope}
\label{sec:pivot}

Any argument that bounds each exterior output's distance to the ideal constant output and then combines two such bounds by the triangle inequality is limited to the saturating family itself. For the dephasing channel at parameter $\varepsilon$, each exterior output of Eq.~\eqref{eq:S_dephasing} sits at trace distance
\begin{equation}
\sqrt{1-|\ip{\chi_{0,1}}{e_0}|^{2}}\;=\;\sqrt{1-(1-\varepsilon)}\;=\;\sqrt{\varepsilon}
\end{equation}
from the ideal pivot $\ketbra{e_0}$, exactly. Such a route, therefore, cannot produce a constant below $2\sqrt{\varepsilon}$, whereas the true pair distance is $2\sqrt{\varepsilon(1-\varepsilon)}<2\sqrt{\varepsilon}$ for every $\varepsilon>0$; the bounds of Sec.~\ref{sec:ksw} are instances of this floor. Reaching the envelope requires analyzing the pair jointly, as the phase-averaged constraint \eqref{eq:S_overlap} does, bounding the pointer \emph{overlap} directly without introducing a third state.

We remark that the restriction to the ideal pivot is essential and not an artifact of the family: for \emph{any} two states, the midpoint $\sigma=\tfrac{1}{2}(\rho+\rho')$ satisfies $\tfrac{1}{2}\lVert\rho-\sigma\rVert_1=\tfrac{1}{4}\lVert\rho-\rho'\rVert_1$ identically, so a pivot chosen with knowledge of the pair reproduces the pair distance exactly, and on the saturating family this midpoint sits at trace distance $\sqrt{\varepsilon(1-\varepsilon)}$ from each extremal output. What no continuity theorem supplies is such a pair-dependent pivot; the pivot they provide is the ideal constant output, and that route is floored at $2\sqrt{\varepsilon}$.

\subsection{Tight trade-offs in a different setting, and the two-state contrast}
\label{sec:dualities}

Tight information-disturbance trade-offs with saturating families are known in a different setting. Hashagen and Wolf \cite{hashagen2019universality} determine optimal trade-offs between measurement error and disturbance, where disturbance is the deviation of the state-evolution channel from the \emph{identity} and the information gained is the quality of approximation to a fixed target measurement. The present setting differs along both axes: our disturbance parameter is the deviation from an \emph{isometry}, and our information side is the distinguishing power of the complementary channel itself, rather than proximity to a chosen measurement. Their optimal tradeoff is attained within a two-parameter family of devices and is largely independent of the chosen distance measures; ours is saturated by dephasing channels realized as which-state measurements whose record resides in the complement. The difference is in the pair of quantities traded off, not in the generality of the analysis. The envelope of Eq.~(1) is not derivable from theirs. The contribution of the present work over this body of results is that envelope, for this pair of quantities, together with its saturation and the which-state equality class that supplies the physical characterization of the extremal horizon.

The fidelity-exact characterization of optimal recovery \cite{beny2010general} expresses the same information-disturbance structure; its approximate form is routed through fidelity-to-trace-distance conversions and therefore carries constants of the type \eqref{eq:S_KSWroute}.

Finally, for two \emph{fixed} orthogonal states, there is no trade-off at all: a nondemolition measurement in their basis distinguishes them perfectly while disturbing neither \cite{fuchs1999cryptographic}. Our hypothesis is worst-case over the sector, including all superpositions of the pair, and this is what forbids the free measurement: the channel that perfectly distinguishes $\ket{0}$ and $\ket{1}$ is complete dephasing, whose fidelity on $\ket{\pm}$ is $\tfrac{1}{2}$, so $\varepsilon=\tfrac{1}{2}$. The global hypothesis is essential to the existence of a nontrivial bound. For energy-constrained continuity in infinite dimensions, see \cite{shirokov2022optimal}; for complementarity of private and correctable structures in infinite dimensions, see \cite{crann2016private}.

\section{Effective interior/exterior factorization}
\label{sec:factorization}

The main text uses an effective bipartition $\mathcal{H}\simeq\mathcal{H}_I\otimes\mathcal{H}_E$. Because a literal factorization does not exist for a sharp horizon, we make precise the sense in which it holds.

\emph{The obstruction.} The algebra of observables of a spacetime subregion is a von Neumann factor of type III$_1$ \cite{witten2018notes}, admitting no tensor factorization of the Hilbert space and no density operators; the Reeh--Schlieder theorem is the physical face of this, forbidding local product states across a sharp horizon.

\emph{Factorization for a stretched horizon.} Let $E_\delta$ denote the exterior up to a stretched horizon, a proper distance $\delta$ outside the true horizon. The split property \cite{doplicher1984standard}, established under standard nuclearity conditions and extended to curved spacetime in \cite{fewster2015split}, guarantees an intermediate type I factor $\mathcal{R}$ with $\mathcal{A}(E_\delta)\subseteq\mathcal{R}\subseteq\mathcal{A}(E)$, and a type I factor induces a genuine tensor factorization. The factorization therefore holds for the quantum-field-theoretic algebras once the horizon is thickened at the Planck scale, subject to the nuclearity hypotheses under which the split property is established \cite{fewster2015split,fewster2016split} and to the proviso that the split property applies to spacelike-separated regions with a finite collar, which is what the stretched horizon supplies; the division across a sharp null horizon is not itself of that form. The edge modes required for a gauge-invariant split \cite{donnelly2016geometric} are the boundary data carried by $\mathcal{R}$.

\emph{The expansion parameter.} In fixed-background quantum field theory, the finite collar regulates the split construction; in gravity, an additional assumption is required that the relevant code subspace admits an approximate interior/exterior organization once the necessary boundary degrees of freedom are included. The remaining approximation is the removal of the collar, $\delta\to 0$. At a short-distance cutoff $a$ the number of independent field cells across the horizon is $N_\partial\sim A/a^{D-2}$, with $D$ the spacetime dimension; at $a\sim\ell_P$ this is $N_\partial\sim A/(4G_N)=S_{\rm BH}$, the $\mathcal{O}(S_{\rm BH})$ boundary-cell count motivates the large-entropy semiclassical description but fixes neither the norm nor the power-law scaling of the factorization error, which we do not derive.

\emph{Gravitational dressing.} Beyond a fixed background the constraints dress subregion observables; adjoining the boundary modular Hamiltonian and an observer converts the type III$_1$ algebra into a type II factor with a trace and density operators, the entropy reducing to the generalized entropy semiclassically \cite{witten2022gravity,chandrasekaran2023large}, and this admits an explicit type I approximation \cite{soni2024type}, constructed for a scalar field on a two-sided black-hole background with a brick wall, which assigns a separable Hilbert space to each exterior for any nonzero wall offset: precisely the stretched-horizon regulator adopted here; the same structure underlies holographic reconstruction as operator-algebra error correction \cite{almheiri2015bulk,pastawski2015holographic}. The reduced states and channels of the main text are well defined in this setting; large $S_{\rm BH}$ motivates the semiclassical expansion, but the cited constructions do not supply a universal norm estimate for the factorization error.

\emph{Scope.} Two physically distinct further departures lie outside the present kinematic setting: the perturbative, constraint-induced availability of information near the boundary, of order $E/M_{\rm Pl}$ for a probe of energy $E$, which is the debate discussed in the main text \cite{laddha2021holographic,raju2022failure,donnelly2018gravitational,bahiru2024holography}; and the $\mathcal{O}(1)$, post-Page-time reorganization of the interior of an evaporating black hole \cite{penington2020entanglement,almheiri2019entropy}, which lies beyond the single-crossing, fixed-background analysis adopted here.

\section{Sector structure and the necessity of the sector restriction}
\label{sec:sectors}

\emph{Classification of the labels.} The background state $\ket{\Phi_0}$ fixes all classical and topological hair already carried by the black hole; distinct values define distinct horizon channels and are not summed over in Eq.~(2). The label $c$ refers only to exterior-accessible conserved or superselected data carried by the infalling system $F$, including any charge whose transfer would shift a background-hair parameter. Exterior observers can determine, without interacting locally with the infalling system, the asymptotic conserved charges $(M,Q,J)$; any \emph{primary} classical hair, that is independent classically measurable exterior parameters such as the non-Abelian data of colored black holes \cite{bizon1990colored,volkov1999gravitating} or scalar charges \cite{herdeiro2015asymptotically} (\emph{secondary} hair, determined by the charges, adds no independent label); and any \emph{superselected} quantum numbers, such as the discrete gauge charges of ~\cite{krauss1989discrete,coleman1992quantum}, which support no classical field and are detectable only by topological Aharonov--Bohm interference. Equation~(2) of the main text is the decomposition over the exterior-accessible labels carried by the infalling system $F$. All classical and topological hair already carried by the black hole remains fixed in $\ket{\Phi_0}$ and labels a different background channel; if $F$ carries a charge whose transfer would change a background-hair parameter, that charge is included in $c$. Hair of the \emph{background} is carried by the fixed state $\ket{\Phi_0}$ and joins $c$ only when the infalling system carries the corresponding charge.

\emph{First prong: the ideal is sector-indexed.} The isometry $V:\mathcal{H}_F^{(c)}\to\mathcal{H}_I$ is fixed by the classical infall trajectory, which itself depends on the sector labels; distinct masses or angular momenta follow distinct trajectories, with distinct backreaction, and define distinct $V$. Assumption (A3) is therefore indexed by the sector at the outset, and no cross-sector hypothesis exists to which the Lemma could be applied. Consistently, were one to posit a single blockwise isometry across such a boundary, any two states $\ket{\psi_c},\ket{\psi_{c'}}$ carrying different labels would be exterior-distinguishable essentially perfectly by asymptotic measurement,
\begin{equation}\label{eq:S_forcing}
\tfrac{1}{2}\bigl\lVert\rho_E(\psi_c)-\rho_E(\psi_{c'})\bigr\rVert_1\;\simeq\;1\,,
\end{equation}
and under such a counterfactual blockwise extension, the smoothness hypothesis would be formulated on their span, so that Eq.~(8) of the main text would force $\varepsilon\geq\tfrac{1}{2}(1-\sqrt{1-D^{2}})\to\tfrac{1}{2}$: such a blockwise extension cannot satisfy the smooth-horizon condition with $\varepsilon<1/2$.

\emph{Second prong: superselected labels.} For labels protected by a genuine superselection rule, such as electric charge or discrete gauge charge, coherent superpositions across sectors are not physical states. The proof of Sec.~\ref{sec:proof} operates on the superpositions $\ket{c_\phi}$, so across such a boundary the hypothesis cannot even be formulated, and such pairs lie outside the domain automatically. Discrete gauge hair thereby converts from an apparent counterexample, being exterior-detectable yet invisible to all classical probes and compatible with a regular horizon, into an illustration of the sector structure: it is a sector label of the second kind.

\emph{Terminology.} We use ``sector'' for the joint refinement of the genuine superselection structure and the classical exterior data. Only the labels of the second prong are superselection sectors in the strict sense; for those of the first, whose cross-sector superpositions are physical, it is the argument \eqref{eq:S_forcing} that operates, and it operates best precisely where superpositions exist. Every exterior-distinguishing label falls under one prong, and the quantity priced by the main text is exclusively the intra-sector distinguishability.

\section{Pre-existing entanglement}
\label{sec:entanglement}

Let $F$ be entangled, before infall, with a reference $R$ that remains outside and never crosses the horizon, the joint input being pure $\ket{\Phi}_{FR}$. The exterior agent controls $(E,R)$, with joint state $\rho_{ER}(\Phi)=(\mathcal{N}_E\otimes\mathrm{id}_R)(\ketbra{\Phi})$; the Stinespring isometry acts as $W\otimes\mathbb{1}_R$.

\emph{Exact case.} At $\varepsilon=0$ the interior channel is exactly isometric, and its minimal dilation acts as $W\ket{\psi}=V\ket{\psi}\otimes\ket{\sigma}$ for a fixed $\ket{\sigma}$, up to an environment isometry that does not affect trace distances (Sec.~\ref{sec:remarks}). Then $(W\otimes\mathbb{1}_R)\ket{\Phi}=[(V\otimes\mathbb{1}_R)\ket{\Phi}]_{IR}\otimes\ket{\sigma}_E$, so $\rho_{ER}(\Phi)=\ketbra{\sigma}\otimes\rho_R(\Phi)$ and, for any two inputs,
\begin{equation}\label{eq:S_exact_ER}
\tfrac{1}{2}\bigl\lVert\rho_{ER}(\Phi)-\rho_{ER}(\Phi')\bigr\rVert_1\;=\;\tfrac{1}{2}\bigl\lVert\rho_R(\Phi)-\rho_R(\Phi')\bigr\rVert_1\,:
\end{equation}
At a perfectly smooth horizon, all exterior distinguishing power resides in the reference. The exterior accesses only correlations that were external to the black hole all along, which is the main text's statement.

\emph{Approximate case.} For $\varepsilon_{\diamond}>0$, the diamond-norm bound of Sec.~\ref{sec:ksw} includes ancillas by definition, so each joint output satisfies $\tfrac{1}{2}\lVert\rho_{ER}(\Phi)-\widetilde{\sigma}\otimes\rho_R(\Phi)\rVert_1\leq\sqrt{2\varepsilon_{\diamond}}$, and the triangle inequality gives
\begin{equation}\label{eq:S_ER_bound}
\tfrac{1}{2}\bigl\lVert\rho_{ER}(\Phi)-\rho_{ER}(\Phi')\bigr\rVert_1\;\leq\;\tfrac{1}{2}\bigl\lVert\rho_R(\Phi)-\rho_R(\Phi')\bigr\rVert_1\;+\;2\sqrt{2\varepsilon_{\diamond}}\,:
\end{equation}
a pre-existing term carried by the reference at no cost in smoothness, plus a crossing-generated term bounded by the horizon disruption. The constants here are those of the general route, Eq.~\eqref{eq:S_KSWroute}; the optimal constant of the Lemma is proven for reference-free same-sector pairs, and it is not established for the entanglement-assisted quantities by the present argument, the obstruction being structural: with a reference present, the ideal joint output is no longer constant, so the pointer analysis does not apply as it stands.
\end{document}